\documentclass[aps,prl,twocolumn,superscriptaddress,reprint,longbibliography]{revtex4-1}

\usepackage{bm}
\usepackage{graphicx}
\usepackage{url}

\usepackage{color}

\def\e0{\varepsilon_0}
\def\ddt{\partial_t}

\def\yv{\hat{\bf y}}
\def\zv{\hat{\bf z}}

\def\bea{\begin{eqnarray}}
\def\eea{\end{eqnarray}}

\begin{document}

\title{Inverse Faraday Effect driven by Radiation Friction}
\author{T. V. Liseykina}\thanks{On leave from Institute of Computational Technologies, SD-RAS, Novosibirsk, Russia}\email{tatyana.liseykina@uni-rostock.de}
\affiliation{Institute of Physics, University of Rostock, 18051 Rostock, Germany}

\author{S. V. Popruzhenko}\email{sergey.popruzhenko@gmail.com}
\affiliation{National Research Nuclear University, Moscow Engineering Physics Institute, Kashirskoe Shosse 31, 115409, Moscow, Russia}

\author{A. Macchi}\email{andrea.macchi@ino.it}
\affiliation{CNR, National Institute of Optics (INO), Adriano Gozzini research unit, Pisa, Italy}
\affiliation{Enrico Fermi Department of Physics, University of Pisa, largo Bruno Pontecorvo 3, 56127 Pisa, Italy}

\date{\today}

\begin{abstract}
A collective, macroscopic signature to detect radiation friction in laser-plasma experiments is proposed. In the interaction of superintense circularly polarized laser pulses with high density targets, the effective dissipation due to radiative losses allows the absorption of electromagnetic angular momentum, which in turn leads to the generation of a quasistatic axial magnetic field. This peculiar ``inverse Faraday effect'' is investigated by analytical modeling and three-dimensional simulations, showing that multi-gigagauss magnetic fields may be generated at laser intensities $>10^{23}~\mbox{W cm}^{-2}$.
\end{abstract}

\maketitle

The development of ultrashort pulse lasers with petawatt power has opened new perspectives for the study of high field physics and ultra-relativistic plasmas \cite{mourouRMP06,dipiazzaRMP12}. In this context, the longstanding problem of radiation friction (RF) or radiation reaction has attracted new interest. 
{RF arises from the back-action on the electron of the electromagnetic (EM) field generated by the electron itself and} plays a dominant role in the dynamics of ultra-relativistic electrons in strong fields. A considerable amount of work has been devoted both to revisiting the RF theory \cite{bulanovPRE11,*kravetsPRE13} and to its implementation in laser-plasma simulations \cite{zhidkovPRL02,sokolovPoP09,tamburiniNJP10,chenPPCF11,vranicXXX15}, {as well as to} the study of radiation-dominated plasmas in high energy astrophysics, see e.g. Refs.\cite{jaroschekPRL09,*ceruttiApJ13,*mahajanMNRAS15}.

{While RF is still an open matter both for classical and quantum electrodynamics \cite{dipiazzaRMP12}}, RF models have not been discriminated experimentally yet. This circumstance led to several proposals of devoted experiments providing clear signatures of RF, e.g. in nonlinear Thomson scattering \cite{kogaPoP05,*dipiazzaPRL09,*hadadPRD10,*neitzPRL13,*blackburnPRL14,*vranicPRL14}, compton scattering \cite{liPRL14}, modification of Raman spectra \cite{kumarPRL13}, electron acceleration in vacuum \cite{tamburiniPRE14,*greenPRL14,*heinzlPRE15}, radiative trapping \cite{gonoskovPRL14,*jiPRL14,*fedotovPRA14} or $\gamma$-ray emission from plasma targets \cite{nakamuraPRL12,*capdessusPRL13}. 
{Most of these studies are based on single particle effects, and RF signatures are found in modifications of observables such as emission patterns and spectra when RF is included in the modeling.} {Detecting such modifications may require substantial improvements in reducing typical uncertainities in laser-plasma experiments.}
{Instead, in this Letter we propose to use a} collective, macroscopic effect induced by RF, namely the generation of multi-gigagauss, quasi-steady, axial magnetic fields in the interaction of a circularly polarized (CP) laser pulse with a dense plasma. This is a peculiar form of the Inverse Faraday effect (IFE) \cite{pitaevskiiJETP61,*pershanPR63,*vanderzielPRL65,*deschampsPRL70} {and may be more accessibile experimentally than single-particle effects. In fact, the IFE has been}  previously studied in different regimes of laser-plasma interactions \cite[and references therein]{steigerPRA72,*abdullaevJETP81,*bychenkhovJETP94,*shengPRE96,*berezhianiPRE97,hainesPRL01,shvetsPRE02}. By using three-dimensional (3D) particle-in-cell (PIC) simulations, we find that at laser intensities foreseeable with next generation facilities producing multi-petawatt \cite{dansonHPL15} or even exawatt pulses \cite{mourouOC12,*mourouNP13}, the magnetic field created by the RF-driven IFE in dense plasma targets reaches multi-gigagauss values with a direction dependent on the laser polarization, which confirms its origin from the ``photon spin''. 
{The magnetic field is slowly varying on times longer than the pulse duration and may be detected via optical polarimetry techniques \cite{horovitzPRL97,*borghesiPRL98,*najmudinPRL01,*tatarakisN02,*wagnerPRE04}, providing an unambiguous signature of the dominance of RF effects.}
{The effect might also be exploited to create strongly magnetized laboratory plasmas in so far unexplored regimes (see e.g. \cite{eliezerPoP05})}.

The IFE is due to absorption of EM angular momentum \footnote{Here we consider only the absorption of intrinsic angular momentum or photon ``spin''. For studies on orbital angular momentum absorption and IFE in laser-plasma interaction see, e.g., Refs.\cite{aliPRL10,*wangSR15}}, 
{which in general is \emph{not} proportional to energy absorption. As an example of direct relevance to the present work, 
let us consider a mirror boosted by the radiation pressure of a CP (with positive helicity, for definiteness) laser pulse. 
From a {quantum-mechanical} point of view, the laser pulse of frequency $\omega$ propagating along $\hat{\bf x}$ corresponds to $N$ incident photons with total energy $N\hbar\omega$ and angular momentum $N\hbar\hat{\bf x}$. If the mirror is perfect, $N$ is conserved in any frame. If the mirror moves along $\hat{\bf x}$, the reflected photons are red-shifted leading to EM energy conversion into mechanical energy (up to 100\% if the mirror velocity $\sim c$) but there is no spin flip for the reflected photons, hence no absorption of angular momentum. However, if the electrons in the mirror emit high-frequency photons, a greater number of incident low-frequency photons must be absorbed with their angular momentum.
From a classical point of view, absorption of angular momentum requires some dissipation mechanism {\cite{shvetsPRE02} which, in our example, implies} a non-vanishing absorption in the rest frame of the mirror.} 

In the case here investigated, effective dissipation is provided by the RF force which makes the electron dynamics consistent with the radiative losses.
{In order to demonstrate IFE induced by RF, we consider a regime of}
ultra-high laser intensity $I_L>10^{23}~\mbox{W cm}^{-2}$ and thick plasma targets (i.e. with thickness much greater than the evanescence length of the laser field) where the {radiative} energy loss is a large fraction of the laser energy {as shown by simulations with RF included} \cite{naumovaPRL09,schlegelPoP09,capdessusPoP14,*capdessusPRE15,nerushPPCF15}. 
{We use a simple model to account for such losses and provide a scaling law with the laser intensity}. The power radiated by an electron moving with velocity $v_x$ along the propagation axis of a CP pulse of amplitude $E_L=(m_e\omega c/e)a_0 \equiv B_0a_0$ (with $\omega$ the laser frequency) is
\begin{eqnarray}
P_{\mbox{\tiny rad}}= \frac{2e^2\omega^2\gamma^2a_0^2}{3c}\left(1-\frac{v_x}{c}\right)^2 \; .
\end{eqnarray}
Since at the relevant frequencies {$\omega_{\mbox{\tiny rad}}\simeq a_0^2\omega$} the radiation is incoherent, the total radiated power by $N$ comoving electrons will be $NP_{\mbox{\tiny rad}}$. For thin targets accelerated by the CP laser pulse (``light sail'' regime), all electrons move with the foil at $v_x \simeq c$, and there is no high-frequency oscillation driven by the ${\bf v}\times{\bf B}$ force. Thus the radiation is strongly suppressed by the factor  $\left(1-{v_x}/{c}\right)^2 \ll 1$, as observed in simulations \cite{tamburiniNJP10,tamburiniPRE12}.  In contrast, RF losses become very important for thick targets \cite{naumovaPRL09,capdessusPRE15,nerushPPCF15} (``hole boring'' regime) because the acceleration of the plasma surface has a pulsed nature \cite{macchiPRL05,schlegelPoP09,schlegelNJP12} with a dense bunch of electrons being periodically dragged towards the incident laser pulse, i.e. in a counterpropagating configuration ($v_x<0$). 

\begin{figure}
\centerline{\includegraphics[width=0.49\textwidth]{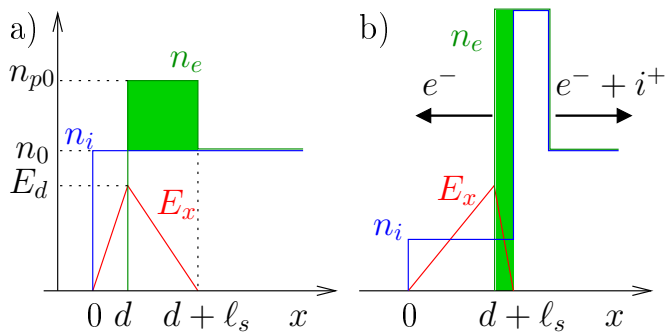}}
\caption{(color online) Cartoon showing the electron dynamics during the ``hole boring'' stage. Frame~a) shows the approximated profiles of the densities of ions ($n_i$) and electrons ($n_e$) and of the electrostatic field ($E_x$) at the early time ($t \simeq 0$) when ions have not moved yet and electrons from the depletion region ($0<x<d$) pile up in the skin layer ($d<x<d+\ell_s$); the number of excess electrons $N_x$ is proportional to the shaded area. Frame~b) corresponds to the time $t \simeq \tau_i$ when the ions have reached the $x\simeq d+\ell_s$ position and formed a quasi-neutral bunch \cite{macchiPRL05}; the excess electrons return towards the depletion region.  
\label{fig:cartoon} 
}
\end{figure}

In order to estimate the number of radiating electrons per unit surface we consider the  dynamic picture of hole boring \cite{macchiPRL05,macchi-book-5.7.2}. As illustrated in Fig.\ref{fig:cartoon}, at the surface of the plasma the radiation pressure generates a positively charged layer of electron depletion (of thickness $d$) and a related pile-up of electrons in the skin layer (of thickness $\ell_s$), i.e. the evanescent laser field region. Ions are accelerated in the skin layer leaving it at a time $\tau_i$ at which an ion bunch neutralized by accompanying electrons is formed. At this instant, the equilibrium between ponderomotive and electrostatic forces is lost and the excess electrons in the skin layer will quickly return back towards the charge depletion region. The number per unit surface of returning electrons is $N_x = (n_{p0}-n_0)\ell_s$ where $n_{p0}$ is the electron density in the skin layer at the beginning of the acceleration stage. Using the model of Refs.\cite{macchiPRL05,macchi-book-5.7.2} $N_x$ may be estimated from the balance of electrostatic and radiation pressures: $e E_d n_{p0}\ell_s/2=2I_L/c$ where $E_d=4\pi e n_0 d$ is the peak field in the depletion region, and $n_{p0}\ell_s=n_0(d+\ell_s)$ because of charge conservation. Solving these equations in the limit $n_{p0}\gg n_0$ we obtain $N_x \simeq {a_0}/{r_c\lambda}$ where $\lambda=2\pi c/\omega$ is the laser wavelength, $r_c=e^2/m_ec^2$ and we used $I_L=m_ec\omega^2a_0^2/(4\pi r_c)$. Thus, the total radiated intensity is $I_{\mbox{\tiny rad}}=P_{\mbox{\tiny rad}}N_x$. In order to compare with the laser intensity $I_L$ we take into account that the radiation is emitted as bursts corresponding to the periodic return of electrons towards the laser, i.e. for a fraction $f_{\tau} \simeq \tau_e/(\tau_e+\tau_i)$ of the interaction stage where $\tau_e$ is the time interval during which the electrons move backwards. Analysis of laser piston oscillations in Ref.\cite{schlegelPoP09} suggests that $\tau_e\simeq \tau_i$ so we take $f_{\tau} \simeq 1/2$ for our rough estimate. Assuming $\left(1-{v_x}/{c}\right)^2 \sim 1$ we obtain for the fraction of radiated energy {to the laser pulse energy}
\begin{equation}
\eta_{\mbox{\tiny rad}} \simeq \frac{4\pi}{3}\frac{r_c}{\lambda}a_0\gamma^2 \; .
\label{eq:eta_rad}
\end{equation}
If the energy of electrons is mainly due to the {motion} in the laser field, then $\gamma \simeq (1+a_0^2)^{1/2} \sim a_0$ for $a_0\gg 1$ and $\eta_{\mbox{\tiny rad}} \propto a_0^3$. For $\lambda=0.8~\mu\mbox{m}$,  $\eta_{\mbox{\tiny rad}} \sim 1$ for $a_0 \sim 400$, corresponding to $I_L \sim 7 \times 10^{23}~\mbox{W cm}^{-2}$. This order-of-magnitude estimate implies that for such intensities a {significant} part of the laser energy is lost as radiation, strongly affecting the interaction dynamics. 
A more precise estimate would require to account both for the energy depletion of the laser and for the trajectory modification of the electrons {due to the RF force}.

A 3D approach is essential to model the phenomena of angular momentum absorption and magnetic field generation, thus we rely on massively parallel PIC simulations {in which RF is implemented following the approach described in Ref.\cite{tamburiniNJP10} (see Ref.\cite{vranicXXX15} for a benchmark with other approaches).} 
{
The laser pulse is initialized in a way that at the waist plane $x=0$ (coincident with the target boundary) the normalized amplitude of the vector potential ${\bf a}=e{\bf A}/m_ec^2$ would be 
\bea
{\bf a}(x=0,r,t) &=&
a_0\left(\yv\cos(\omega t)\pm \zv\sin(\omega t)\right)
\nonumber\\
& &\times 
{
\mbox{e}^{-(r/r_0)^n-(ct/r_l)^4}
} \; ,
\label{eq:laserpulse}
\eea
where $r=(y^2+z^2)^{1/2}$. 
{Both {radial} profiles with $n=2$ (Gaussian, G) and $n=4$ (super-Gaussian, SG) have been used in the simulations.}
For all the results shown below, we take $r_l=3\lambda$ and radius $r_0=3.8\lambda$. The plus and minus sign in the expression for ${\bf a}$ correspond to
positive and negative helicity, respectively. 
The pulse energy is given by $U_L={\cal A} {r_0^2r_l} B_0^2a_0^2$ where {${\cal A}=\Gamma({1}/{4}){2^{-17/2}} \simeq 0.19$} and {${\cal A}={\pi^{1/2}\Gamma({1}/{4})}{2^{-19/2}} \simeq 0.24$} {for the G and SG pulse cases, respectively.} The target is a plasma of thickness $10\lambda$ and electron density $n_0=90n_c$ (where $n_c=m_e\omega^2/(4\pi e^2)$ is the cut-off density) and charge-to-mass ratio for ions $Z/A=1/2$. The range of laser amplitudes investigated in the simulation is $a_0=200-600$.
Assuming $\lambda=0.8\mu\mbox{m}$, the density $n_0=1.55 \times 10^{23}~\mbox{cm}^{-3}$, the pulse duration  (full-width-half-maximum of the intensity profile) is $14.6$~fs and the range for the peak laser intensity $I_L=m_ec^3n_ca_0^2$ is $(1.9-16.7) \times 10^{23}~\mbox{W cm}^{-2}$ corresponding to a pulse energy {$U_L\simeq(0.38-3.4)$~kJ for the G pulse} and {$U_L\simeq(0.48-4.3)$~kJ} for the SG pulse. The numerical box had a $30\times25\times 25\lambda^3$ size, with 40 grid cells per $\lambda$ and 64 particles per cell for each species.
The simulations were performed on 480 cores of the JURECA supercomputer at Forschungszentrum J\"ulich. 
}

\begin{figure}
\centerline{\includegraphics[width=0.5\textwidth]{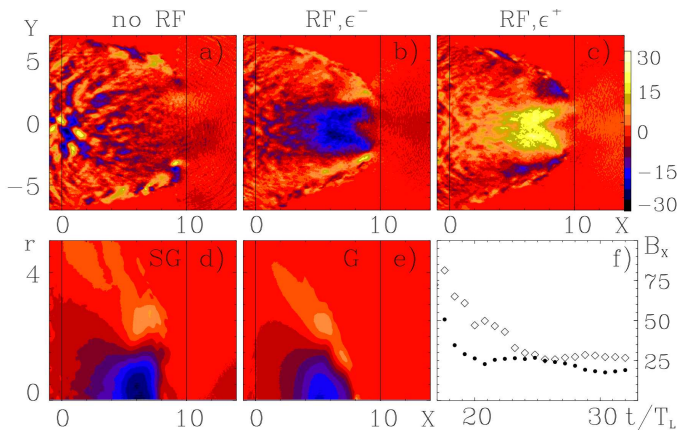}}
\caption{(color online) a)-c): Axial magnetic field $B_x$ (normalized to $B_0=1.34 \times 10^8$~G) in 3D simulation with a super-Gaussian pulse. Case~a) is without RF, case b)~c) are with RF included and for opposite {helicities}. 
The field is shown in the $xy$ plane {$t=27\lambda/c$} after the beginning of the interaction (very similar patterns are observed in the $xz$ plane, not shown). The laser pulse is incident along the $x$ axis from the left side and the thin black lines denote the boundaries of the target. The coordinates are normalized to $\lambda$. d) and e): $B_x$ averaged over the azimuthal direction comparison for Gaussian (G) and super-Gaussian (SG) pulse profiles. f): the temporal evolution of the maximum value of $B_x$ on the $x$ axis for both the G (filled dots) and SG (empty diamonds) pulses. {The time $t=0$ corresponds to the laser pulse peak reaching the waist, as in Eq.(\ref{eq:laserpulse}).} \label{fig:Bx} 
}
\end{figure}

Figures~\ref{fig:Bx}~a)-c) show the magnetic field $B_x$ (normalized to $B_0=1.34 \times 10^8$~G for $\lambda=0.8~\mu\mbox{m}$) along the propagation direction at time $t=27\lambda/c$ for a simulation [a)] where RF is not included and for two simulations [b)-c)] including RF and having positive and negative helicity, respectively; the laser profile was super-Gaussian and $a_0=600$ in all the three simulations. %
Only with RF included an axial magnetic field of maximum amplitude $B_{\mbox{\tiny max}}\simeq 22B_0=2.9 \times 10^9$~G, extending over several microns and a polarity inverting with the pulse helicity is generated.
{The comparison of Fig.\ref{fig:Bx}~d) and e) shows that $B_x$ has similar values and extension for a Gaussian pulse. The field is slowly varying over more than a ten laser cycles ($\sim$30~fs) time, with no sign of rapid decay at the end of the simulation, as shown in Fig.\ref{fig:Bx}~f).
}

\begin{figure}
\includegraphics[width=0.49\textwidth]{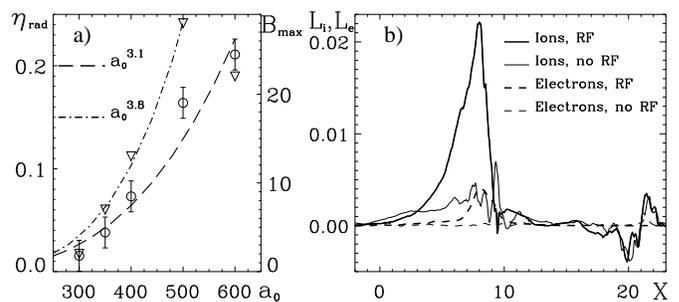}
\caption{a) Values of fractional radiative energy loss $\eta_{\mbox{\tiny rad}}$ (circles) and maximum axial magnetic field $B_{\mbox{\tiny max}}$ (triangles) as a function of laser amplitude $a_0$, from 3D simulations. The dashed and dash-dotted lines are fit to the data for $\eta_{\mbox{\tiny rad}}$ and $B_{\mbox{\tiny max}}$, respectively. The errorbars on $\eta_{\mbox{\tiny rad}}$ correspond to the typical $\lesssim 1\%$ amount of energy which is not conserved in the simulation because of numerical errors.\label{fig:scaling}
b) Axial angular momentum of electrons (dashed line) and ions (thick line) at $t=27\lambda/c$. The density of angular momentum has been integrated over the radius {and normalized to the total integrated angular momentum of the laser pulse.} Results with and without RF included are shown.\label{fig:amarpa}
}
\end{figure}

The fraction $\eta_{\mbox{\tiny rad}}$ of the laser energy dissipated by RF reaches values up to $\eta_{\mbox{\tiny rad}} \simeq 0.24$ for $a_0=600$ as shown in Fig.\ref{fig:scaling}~a) .
A fit to the data gives $\eta_{\mbox{\tiny rad}}\propto a_0^{3.1}$, close to the $\eta_{\mbox{\tiny rad}} \sim a_0^3$ prediction of our model. Fig.\ref{fig:scaling}~a) also shows the peak magnetic field $B_{\mbox{\tiny max}}$ {scaling as $\sim a_0^{3.8}$} up to the highest value $B_{\mbox{\tiny max}}\simeq 28B_0=3.75~\mbox{GG}$ for $a_0=500$. The decrease down to $B_{\mbox{\tiny max}}\simeq 22B_0$ for $a_0=600$ is related to the early interruption of the hole boring stage due to the breakthrough of the laser pulse through the target as observed in this case.
{Notice that we do not show simulations for $a_0<200$ since in such case the RF losses become too close to the percentage of energy which is lost due to numerical errors ($\lesssim 1\%$). However, the inferred scaling would predict $B_{\mbox{\tiny max}} \sim 8~\mbox{MG}$ for $a_0=100$, {which may be still detectable making an experimental test closer}}. 
{To sketch an analytical model for IFE, let us first observe that the density of angular momentum of the laser pulse 
{${\cal L}_z={\bf r}\times({\bf E}\times{\bf B})/4\pi c=-r\partial_rI_L(r)/(2c\omega)$}, 
with $I_L(r)$ the radial profile of the intensity, vanishes on axis and has its maximum at the edge of the beam. We thus consider angular momentum absorption to occur in a thin cylindrical shell of radius $R \simeq r_0$, thickness $\delta\ll R$, and length $h$. The temporal growth of the axial field $B_x$ induces an azimuthal electric field $E_{\phi}$, which in turn allows the absorbed angular momentum to be transfered from electrons to ions. 
{Assuming that the electron and ion shells rotate with angular velocities $\Omega_{e,i}$, respectively, we may write for the angular momenta $L_e={\cal I}_e{\Omega_e}$ and $L_i={\cal I}_i{\Omega_i}$ {where ${\cal I}_e=2\pi R^3\delta h m_e n_e$ and ${\cal I}_i=(Am_p/Zm_e){\cal I}_e$ are the momenta of inertia for electrons and ions, respectively}. The global evolution of the angular momenta of electrons and ions may be described by the equations
\begin{equation}
{\cal I}_e\frac{d\Omega_e}{dt}=M_{\mbox{\tiny abs}}-M_E \; , 
\qquad 
{\cal I}_i\frac{d\Omega_i}{dt}=M_E \; ,
\label{eq:LeLi}
\end{equation}
where $M_{\mbox{\tiny abs}}$ is the torque due to angular momentum absorption (related to the absorbed power $P_{\mbox{\tiny abs}}$ by $M_{\mbox{\tiny abs}}=P_{\mbox{\tiny abs}}/\omega$) and $M_E$ is the torque due to $E_{\phi}$:
\bea
M_E =\int eE_{\phi}(r)rn_ed^3r \simeq \frac{eE_{\phi}(R)}{m_eR}{\cal I}_e \; . 
\label{eq:ME}
\eea
The rotation of the electrons induces a current density $j_{e\phi} \simeq -en_e\Omega_e R$. Neglecting the displacement current, in the limiting case $h \gg R$ the field $B_x \simeq 4\pi j_{e\phi}\delta/c$ and it is uniform as in a solenoid. In the opposite limit $h \sim \delta \ll R$, the current distribution may be approximated by a thin wire of cross-section $\sim h\delta$, and $E_{\phi}(R)$ can be obtained via the self-induction coefficient of a coil \cite{jackson-Mcoil}. We thus obtain
\bea
M_E \simeq 
{\cal F} \frac{\omega_p^2R\delta}{2c^2}{\cal I}_e\frac{d\Omega_e}{dt}
\equiv {\cal I}'_e\frac{d\Omega_e}{dt} \; ,
\eea
where $\omega_p^2=4\pi n_ee^2/m_e$. The geometrical factor ${\cal F} \simeq 1$ if $h\gg R$, and ${\cal F} \simeq (h/R)\ln(8R/\sqrt{h\delta})$ if $h \simeq \delta \ll R$. We thus obtain
\bea
\Omega_e(t)=\frac{1}{{\cal I}_e+{\cal I}'_e}\int_0^tM_{\mbox{\tiny abs}}(t')dt' \; , 
\eea
which shows that the electron rotation follows promptly the temporal profile of $M_{\mbox{\tiny abs}}(t)$, and that effect of the inductive field on electrons is equivalent to effective inertia. Since in our conditions ${\cal I}'_e \sim (\omega_p^2/\omega^2){\cal I}_e=(n_e/n_c){\cal I}_e \gg {\cal I}_e$, the l.h.s. term in Eq.(\ref{eq:LeLi}) can be neglected and $M_E \simeq M_{\mbox{\tiny abs}}$ holds. Thus, from Eq.(\ref{eq:LeLi}) we obtain }
\bea
L_i \simeq \int_0^{t}M_{\mbox{\tiny abs}}(t')dt'\simeq \frac{{\cal I}'_e}{{\cal I}_e}L_e \gg L_e \; ,
\eea
}
{i.e.} the total angular momentum of ions is much larger that of electrons. This is in agreement with the simulation results [Fig.\ref{fig:amarpa}~b)].

{In turn, posing $M_E \simeq M_{\mbox{\tiny abs}}$ in Eq.(\ref{eq:ME}) and using $E_{\phi}(R)\simeq -(R/2c){\cal G}\ddt B_x(r=0,t)$ [where ${\cal G}=1$ for $h\gg R$ and ${\cal G}\simeq (2/\pi)\ln(8R/\sqrt{h\delta})$ for $h \sim\delta\ll R$] we obtain for the final value of the magnetic field on axis $B_{xm}=B_x(r=0,t=\infty)$
\bea
\frac{\pi e}{c}n_eh R^3\delta{\cal G} B_{xm} \simeq \int_0^{\infty}M_{\mbox{\tiny abs}}(t)dt =L_{\mbox{\tiny abs}} \; . 
\eea
The total angular momentum absorbed $L_{\mbox{\tiny abs}}=U_{\mbox{\tiny abs}}/\omega$ where the absorbed energy is $U_{\mbox{\tiny abs}} \simeq \eta_{\mbox{\tiny rad}}U_L$, assuming RF as the main source of dissipation. 
We thus estimate the final magnetic field as
\bea
\frac{B_{xm}}{B_0} \simeq \frac{{\cal A}}{\pi{\cal G}}\frac{\eta_{\mbox{\tiny rad}}}{n_eh}\frac{B_0c}{e\omega}\frac{r_l}{R\delta}a_0^2 \; .
\eea
}

The product {$n_eh$} is the surface density of the region where dissipation and angular momentum absorption occur. Thus, with reference to Fig.\ref{fig:cartoon} we may estimate $n_eh \simeq n_{p0}\ell_s \simeq (I_L/\pi e^2c)^{1/2}=2 n_ca_0c/\omega$ (for $n_{p0}\gg n_0$). Noticing that $B_0/en_c=2\lambda$ we eventually obtain
\begin{eqnarray}
\frac{B_{xm}}{B_0} \simeq 
\frac{{\cal A}}{\pi{\cal G}}\eta_{\mbox{\tiny rad}}\frac{r_l\lambda}{R\delta}a_0\; .
\label{eq:Bxs}
\end{eqnarray}
If $\eta_{\mbox{\tiny rad}} \propto a_0^3$ then $B_{xm} \propto a_0^4$, in good agreement with the observed scaling in Fig.\ref{fig:scaling}~a). 
{If we pose $R \simeq r_0$, the laser initial beam radius, and $\delta \simeq \lambda$, the radial width of the angular momentum density}, for $a_0=500$, $\eta_{\mbox{\tiny rad}}=0.16$ {and ${\cal G}=1$} Eq.(\ref{eq:Bxs}) yields $B_{xm} \simeq 4.8B_0$. The discrepancy with the observed value of $\simeq 28B_0$ may be {attributed to the nonlinear evolution and self-channeling of the laser pulse in the course of the hole boring process. For instance, Fig.\ref{fig:Bx} shows that the magnetic field is generated in a region of radius $\sim 2\lambda$. Further analysis of the simulation data shows both a slight increase (by a factor $\sim 1.2$) of the laser amplitude on the axis and a localizaton of the densities of both EM and mechanical angular momenta in a narrow layer of $\sim 0.5\lambda$ width. Posing $R\simeq 2\lambda$, $\delta\simeq 0.5\lambda$ and an effective $a_0 \simeq 600$ in the above estimate yields $B_{xm} \simeq 23B_0$, which is in fair agreement with the simulation results} considering the roughness of the model. 

In conclusion, we showed in 3D simulations that in the interaction of superintense, circularly polarized laser pulses with thick, high density targets the strong radiation friction effects lead to angular momentum absorption and generation of {multi-}gigagauss magnetic fields via the Inverse Faraday effect. Simple models for the efficiency of radiative losses, the transfer of angular momentum to ions and the value of the magnetic field are in fair agreement with the simulation results for what concerns both the scaling with intensity and order-of-magnitude estimates.  
With the advent of multi-petawatt laser systems, the investigated effect may {provide a laboratory example of radiation-dominated, strongly magnetized plasmas and} a macroscopic signature of radiation friction, {providing a test bed for related theories.}

\begin{acknowledgments}
Suggestions from D.~Bauer are gratefully acknowledged.
The simulations were performed using the computing resources granted by the John von Neumann-Institut f\"ur Computing (Research Center J\"ulich) under the project HRO01. T.V.L.  acknowledges DFG within the SFB 652. S.P. acknowledges support of the excellence center for applied mathematics and theoretical physics within MEPhI Academic Excellence Project (contract No. 02.a03.21.0005, 27.08.2013). 
\end{acknowledgments}


\hyphenation{Post-Script Sprin-ger}

\end{document}